# A frequency-tunable nanomembrane mechanical oscillator with embedded quantum dots

*Xueyong Yuan[1,2], Michael Schwendtner[1], Rinaldo Trotta[1,3], Yongheng Huo[1,4,5]\*, Javier Martín-Sánchez[1,6,7]\*, Giovanni Piredda[8], Huiying Huang[1], Johannes Edlinger[8], Christian Diskus[9], Oliver G. Schmidt[4], Bernhard Jakoby[9], Hubert J. Krenner[2,10], Armando Rastelli[1]\**

[1] Institute of Semiconductor and Solid State Physics, Johannes Kepler University Linz, Altenbergerstraße 69, 4040, Linz, Austria

[2] Lehrstuhl für Experimentalphysik 1 and Augsburg Centre for Innovative Technologies (ACIT), Universität Augsburg, Universitätsstraße 1, 86159 Augsburg, Germany

[3] Department of Physics, Sapienza University of Rome, Piazzale Aldo Moro 5, 00185, Rome, Italy

[4] Institute for Integrative Nanosciences, IFW Dresden, Helmholtzstraße 20, 01069 Dresden, Germany

[5] Hefei National Laboratory for Physical Sciences at Microscale, University of Science and Technology of China, Shanghai Branch, Xiupu Road 99, Shanghai, 201315, China

[6] Departamento de Física, Universidad de Oviedo, 33007, Oviedo, Spain

[7] Centro de Investigación en Nanomateriales y Nanotecnología, CINN (CSIC—Universidad de Oviedo), El Entrego 33940, Spain

[8] Forschungszentrum Mikrotechnik, FH Vorarlberg, Hochschulstraße 1, 6850 Dornbirn, Austria

[9] Institute for Microelectronics and Microsensors, Johannes Kepler University Linz, Altenbergerstraße 69, 4040, Linz, Austria

[10] Nanosystems Initiative Munich (NIM), Schellingstraße. 4, 80799 München, Germany

\*corresponding authors: yongheng@ustc.edu.cn; javiermartin@uniovi.es; armando.rastelli@jku.at

**Abstract**

Hybrid systems consisting of a quantum emitter coupled to a mechanical oscillator are receiving increasing attention for fundamental science and potential applications in quantum technologies. In contrast to most of the presented works, in which the oscillator eigenfrequencies are irreversibly determined by the fabrication process, we present here a simple approach to obtain frequency-tunable mechanical resonators based on suspended nanomembranes. The method relies on a micromachined piezoelectric actuator, which we use both to drive resonant oscillations of a suspended Ga(Al)As membrane with embedded quantum dots and to fine tune their mechanical eigenfrequencies. Specifically, we excite oscillations with frequencies of at least 60 MHz by applying an AC voltage to the actuator and tune the eigenfrequencies by at least 25 times their linewidth by continuously varying the elastic stress state in the membranes through a DC voltage. The light emitted by optically excited quantum dots is used as sensitive local strain gauge to monitor the oscillation frequency and amplitude. We expect that our method has the potential to be applicable to other optomechanical systems based on dielectric and semiconductor membranes possibly operating in the quantum regime.

Micro- and nano-mechanical resonators are widely employed as sensors because of their response to, e.g., electric, magnetic and optical forces. New avenues for nano-mechanical resonators have recently opened up thanks to the demonstrated capability of reaching the vibrational ground state of such systems[1], of achieving quantum entanglement among remote optomechanical oscillators[2], as well as from the combination of such mesoscopic objects with two-level quantum systems[3,4]. Ideas range from mechanical control of quantum mechanical systems such as single spins[5], quantum non-demolition (QND) measurements of the state of the quantum systems by reading the mechanical state of the resonator[6], to the use of mechanical oscillations to mediate the interaction between distinct quantum mechanical systems.[7,8]

For some applications relying on multiple resonators, such as in Ref.[2], it would be useful to have oscillators featuring the same eigenfrequencies, which is particularly challenging for resonances with high quality (Q) factors, due to unavoidable fluctuations in the fabrication process. The eigenfrequencies of an oscillator are determined not only by the geometrical parameters and mass density, but can be controlled also after fabrication[9]. Fine tuning of the mechanical resonances has been achieved by changing the oscillator temperature[10] – an approach poorly suited for oscillators operated in the quantum regime – and by changing the strain state of the oscillator, either by electrostatic[11–13] or by thin-film piezoelectric[14] actuation. The basic physical principle behind frequency tuning is that mechanical tension leads to a stiffening of the material and thus to an increase of the resonance frequencies[15], as well known from guitar strings.

Here we demonstrate frequency tuning by employing a micromachined piezoelectric actuator[16] capable of inducing large and reversible uniaxial stress on an overlying membrane, with strain values exceeding 1% also at cryogenic temperatures. For a proof of principle, we employ doubly-clamped beam oscillators with mass of a few hundred picograms made of Ga(Al)As, which are directly bonded onto the actuator. The choice of this material class is motivated by their compatibility with complex optomechanical-circuit fabrication[17] and with monolithic integration of high-quality quantum QDs[3,18,19]. Specifically, we use single InGaAs and GaAs quantum dots (QDs) to monitor the vibrations. We also show that the actuator can be easily used to drive oscillations with frequencies of at least 60 MHz, which is already more than one order of magnitude larger than the frequencies explored so far in QD-nanomechanical systems[18–20]. (Higher vibrational frequencies have only been achieved for propagating nanomechanical waves[21–23]).

The used piezoelectric actuators were derived from 500-µm thick (1-$x$)[Pb(Mg$_{1/3}$Nb$_{2/3}$)O$_3$]-$x$[PbTiO$_3$] (PMN-PT) plates, which were mechanically lapped and polished down to a final thickness of ~300 µm, microprocessed via a femtosecond laser (see Ref.[24]) to feature two finger structures, and metallized, as described in Ref.[16].

The semiconductor samples were grown by molecular beam epitaxy (MBE) on GaAs(001) substrates. The optically active layers (with thickness of ~250 nm) containing the QDs were grown on a 100-nm thick Al$_{0.7}$Ga$_{0.3}$As sacrificial layer, which can be removed by selective etching, allowing the nanomembranes to be easily released from the substrate. The InGaAs QDs were obtained via the Stranski-Krastanow (SK) growth mode and the GaAs QDs with the local droplet etching method[25,26] and were placed at the center of the membrane along the growth direction. After growth, standard optical photolithography, metallization with Cr/Au (3/100 nm), and chemical etching were performed to obtain stripe-shaped nanomembranes with typical size of 4×300 µm$^2$. The metal layer is used as lithography mask and to displace the QD layer with respect to the whole membrane center (neutral plane), so that the QDs are subject to strain both for longitudinal and transversal oscillations, as discussed later. (The membranes structures can be found in Fig. S1 of Supplementary Material). The long edges of the membranes are parallel to the [100] crystal direction of GaAs to avoid the occurrence of piezoelectric polarization when uniaxial stress is applied for frequency tuning. After that, the nanomembranes were bonded onto the PMN-PT actuator via a flip-chip process, with the stripes placed on the gap between the finger structures. SU8 photoresist was used as a glue for its strong adhesiveness at relatively low temperature[27]. Substrate removal via wet chemical etching leaves suspended membranes, which we use as frequency-tunable micromechanical oscillators, see Fig.1. The device was finally placed on an AlN chip carrier, which provides both good electric insulation and thermal conduction for operation at cryogenic temperature in a He-flow cryostat. For more detailed information about the device process, see Refs.[16,28,29].

With the PMN-PT actuator, we not only excite mechanical oscillations (as in Refs.[18,20]), but also tune the intrinsic stress of the nanomembranes (as in Ref.[16]). To this end, a voltage $V(t)=V_{DC}+V_{AC}(t)$ is applied to the bottom sides of the actuator fingers with respect to the top face (grounded) via a bias-tee, inducing a contraction (for V>0) or expansion (for V<0) of the fingers and thus tensile or compressive strain in the suspended membrane, respectively (see Fig. 1). Since the total length of the fingers (3 mm) is much larger than the gap width (20-50 µm), the strain in the membrane is geometrically enlarged compared to the strain in the PMN-PT[16].

To monitor the oscillations, we use the strong coupling between strain in the membrane oscillator and the optical emission of the embedded QDs[18,20,30]. To this end, we excite photoluminescence (PL) from single QDs at a sample temperature of ~8 K using a continuous-wave (cw) laser with wavelength of 532 nm, focused by a 50× objective with numerical aperture of 0.42. The PL signal emitted from QDs is collected by the same objective and detected with a liquid-nitrogen-cooled silicon Charge-Coupled-Device (CCD) after being dispersed by a spectrometer equipped with 1800 *grooves/mm* grating.

When the nanomembrane runs into a transverse vibrational resonance, mechanical oscillations occur, in which the QDs will be alternatively subject to compressive and tensile strain (illustrated by the finite-element-method simulations in Fig. 2 (a)). By recording PL spectra while scanning the frequency $f$ of the driving signal $V_{AC}(t)=V_0 \sin(2\pi f t)$, we detect the occurrence of mechanical resonances as a significant strain-induced line broadening, as shown in Figs. 2(b, c) for the neutral exciton (X) emission of an InGaAs QD. The observed lineshape stems from the periodic oscillations of the line position, which are averaged over the PL acquisition time (~1 s, which is much larger that the oscillation period $1/f$). Assuming a Lorentzian lineshape for the unperturbed QD emission, the time-integrated spectrum $I(E)$ (PL intensity vs emission energy $E$) for a driving frequency $f$ can be fit with[31,32]

$$I = I_0 + f\frac{2A}{\pi}\int_0^{1/f}\frac{w}{4\times(E-(E_0+\Delta E\times\sin(2\pi\cdot f\cdot t)))^2+w^2}dt, \qquad (1)$$

where $I_0$, $A$, $w$, $E_0$, and $\Delta E$ are, respectively, a constant offset, the line intensity, linewidth, central energy, and strain-induced energy shift. Fig. 2(b) shows the fit of the emission broadening under resonant frequencies $f_0$ = 1.55 MHz and 9.19 MHz. Using equation (1), we extract spectral broadenings of $2\Delta E$ = 622 μeV and 215 μeV for the two resonances, respectively. Fig. 2(c) shows the time-averaged PL emission as the frequency $f$ is swept from 1 MHz to 10 MHz, with an AC voltage amplitude $V_0$=2.0 V and no external axial stress applied ($V_{DC}$=0). The corresponding extracted spectral broadening $2\Delta E$ as a function of frequency $f$ is plotted in Fig. 2(d). At least five sharp resonant peaks (marked as $r_i$ (i=1-5) in Fig. 2(c,d) ) are clearly observable in this range, which indicate distinct mechanical resonances of the suspended nanomembrane.

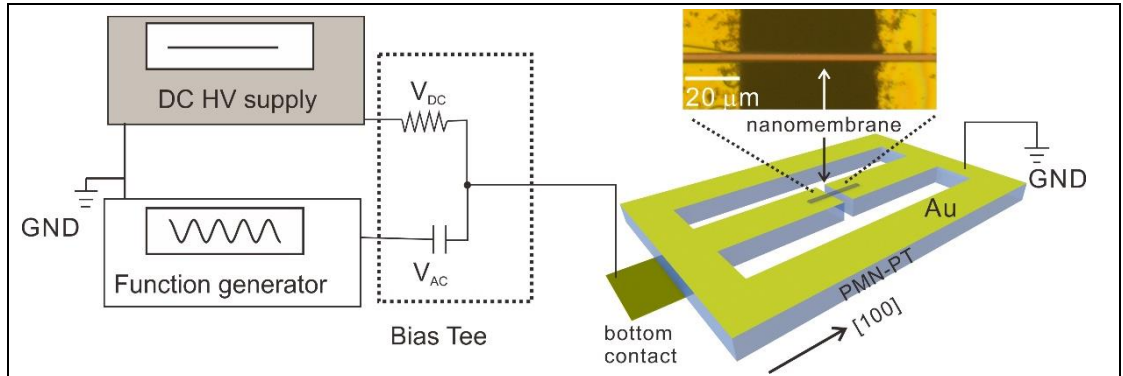

Figure 1. Schematic picture of the device used in this work. The right panel shows a sketch of the micromachined PMN-PT actuator, which we use to both drive oscillations in a mechanical resonator and tune their eigenfrequencies. The inset shows an optical microscopy picture of a suspended nanomembrane oscillator with embedded QDs. For a proof of principle, the oscillator is directly bonded on the two fingers of the PMN-PT actuator. A DC high-voltage (HV) supply and a function generator are employed to apply DC and AC signals to the bottom contacts of the PMN-PT actuator (extending under the two fingers), while the top face is grounded.

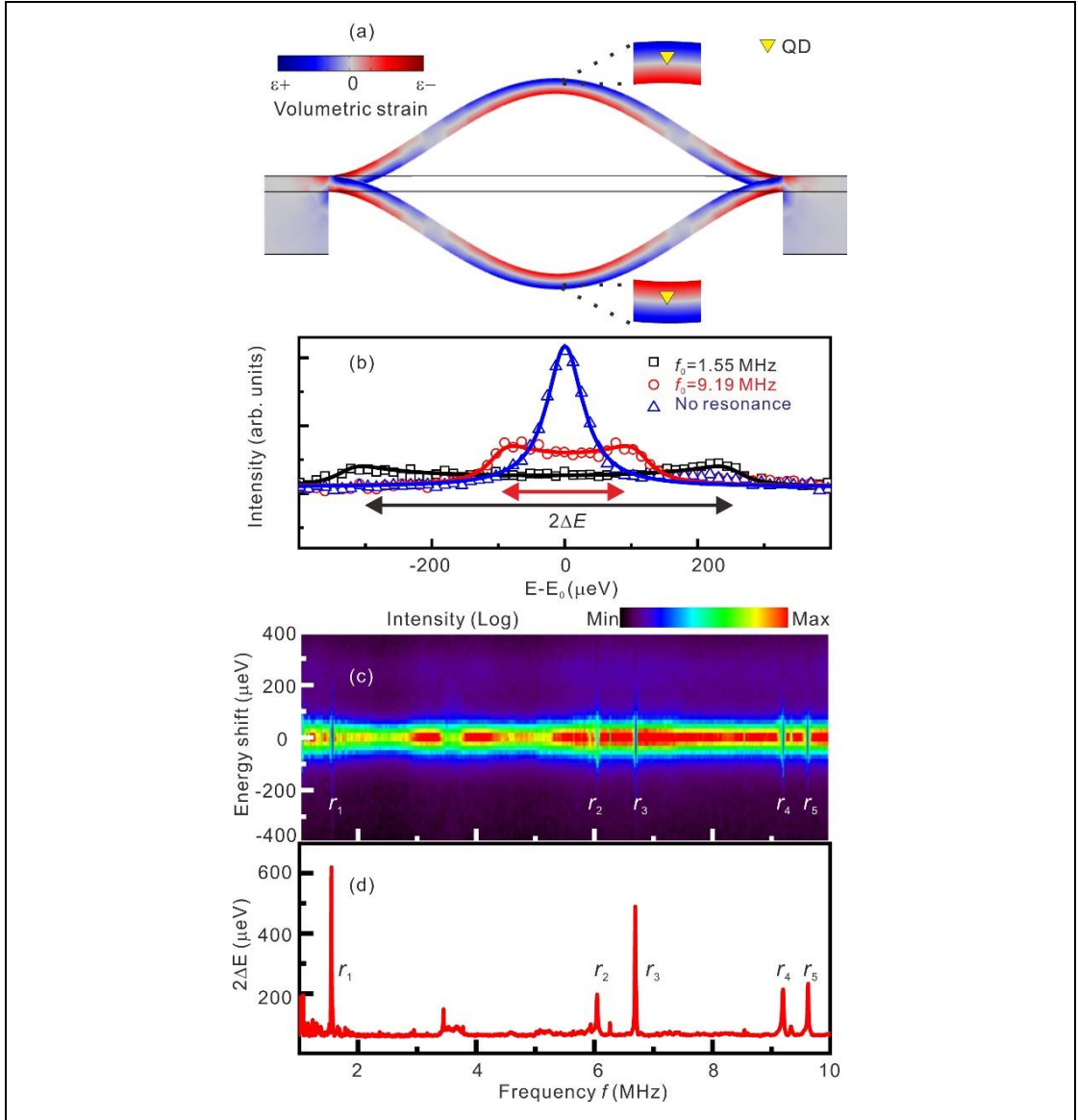

Figure 2. (a) Sketch of the nanomembrane deformation for the first-order transversal vibration mode. The strain field, calculated by the finite-element-method, is plotted in blue-to-red scale (red: compression, blue: tension). Due to the off-center position, a QD (indicated by inverted yellow triangles) experiences strain fields, which we detect as emission energy shifts. (b) PL emission of an exciton confined in an InGaAs QD under sinusoidal driving at resonant frequencies $f_0$ = 1.55 MHz, 9.19 MHz, and out of resonance. Solid lines are fits based on Eq. (1), while the off-resonance emission is fitted with a Lorentzian. (c) Color-coded plot of the time-integrated emission of the same QD as a function of the driving frequency $f$ between 1 and 10 MHz, with AC driving voltage $V_0$ = 2.0 V and $V_{DC}$ = 0 V. (d) Extracted spectral broadening $2\Delta E$ dependence on the driving frequency $f$. $r_i$ (i=1,2…5) indicate some of the clearly resolved resonance peaks.

We now turn to the frequency tuning of the mechanical resonances of our simple beam oscillator. Fig. 3(a) shows one of the resonances and its evolution with varying driving amplitude $V_0$. Here a trion emission line of an InGaAs QD is chosen because of its relatively narrow linewidth ($w = 29$ μeV, corresponding to the resolution limit of the used setup) and relatively high intensity. An example of data used to generate Fig. 3(a) is shown in the inset, which presents color-coded PL spectra collected for different driving frequencies with a fixed driving amplitude $V_0 = 100$ mV. At low driving amplitude, the lineshape of the resonance is symmetric and, for $V_{DC} = 0$ V, it has a quality factor of $5446 \pm 2$.

With increasing driving amplitudes, the peak becomes asymmetric and shifts to lower frequency. This "shark-fin shape" can be attributed to the onset of nonlinear behavior, which can be modelled with a Duffing oscillator model[33]. Although the origin of the non-linearity is currently not clear, we have observed a similar behavior in different devices, consistent with previous investigations in other nanomechanical oscillators[34,35]. In order to exclude non-linear effects and investigate the tuning of the resonance frequency, a smaller driving amplitude of 25 mV is used, which is sufficient to evaluate the tuning behavior when a variable uniaxial stress is introduced by varying $V_{DC}$. The resonance evolution with different DC voltages (tensile stresses) is plotted in Fig. 3(b). With increasing static tensile stress, the PL emission shows a monotonic redshift, similar to the GaAs QDs studied in Ref.[16]. At the same time, the resonance frequency (marked by red dashed circles) increases with increasing tensile stress, as expected from previous works relying on static stress[15]. This phenomenon can be attributed to a stiffening of the material under tension, which is well known for chord music instruments. Although the frequency tuning shown in Fig. 3(b-c) is not large (~0.46%, from 2.772 MHz to 2.785 MHz) because of the limited applied $V_{DC}$ (uniaxial strain around 0.03%), the shift is already ~25 times larger than the resonance linewidth. The frequency shift is accompanied by a marked increase of Q factor from 5446 to 7284 (shown in Fig. 3(c)), qualitatively consistent with previous results obtained with static tension[36–38]. The reason for the limited tuning range shown here is that often resonances disappeared during the experiment, probably due to irreversible structural changes (for instance a change of the bonding layer or detachment of particles present on the membrane). In another case, the resonance frequency was changed by as much as 8% of its original value (see Supplementary Material Fig. S2). In this case, the estimated strain value is about 0.35%.

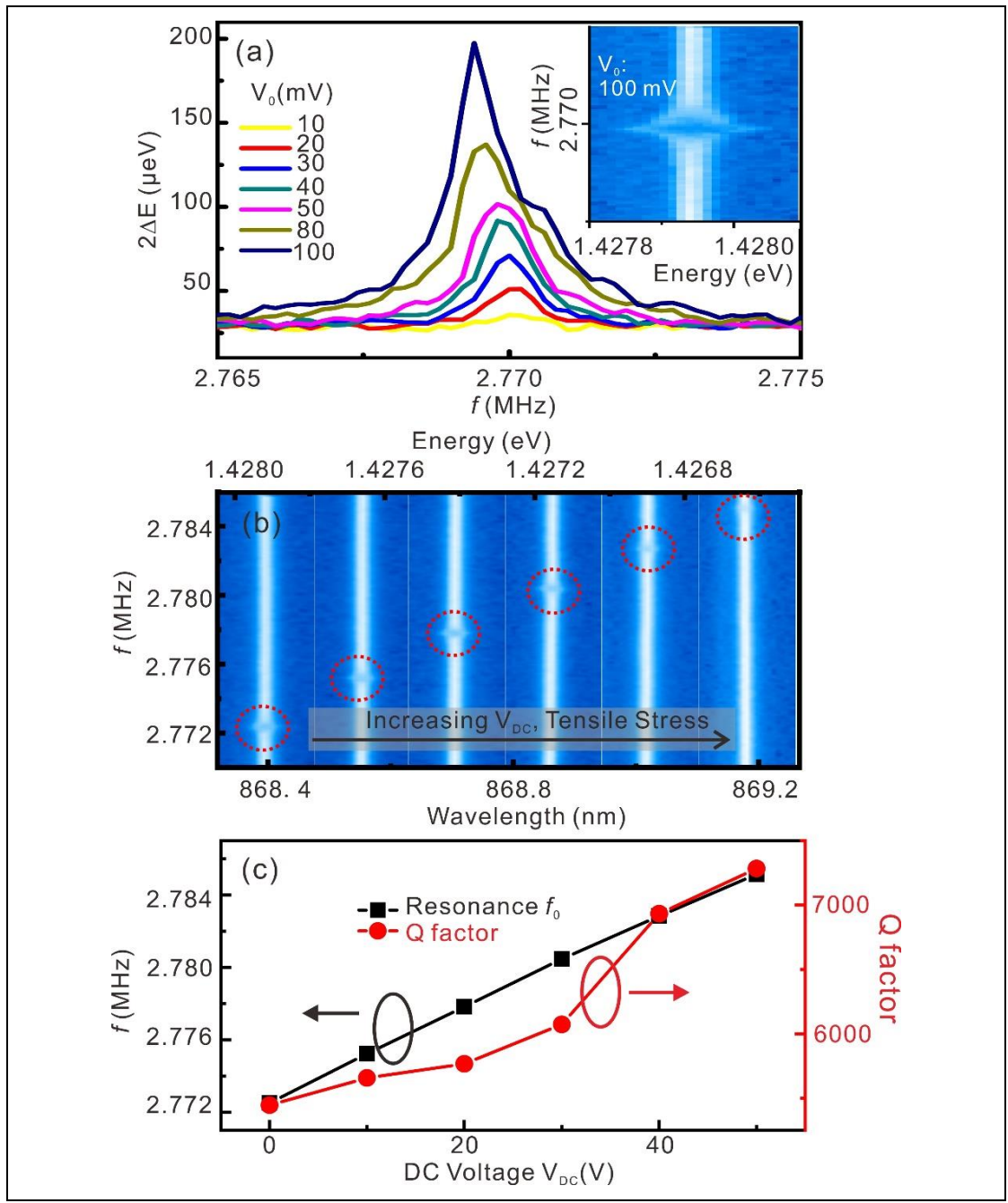

Figure 3. (a) Variations of the spectral broadening 2ΔE of the trion emission peak from an InGaAs QD with different sinusoidal driving amplitudes at a resonant frequency $f_0$=2.770 MHz. The DC voltage was kept at 0 V (no external static axial stress is applied through the PMN-PT actuator). The inset shows color-coded PL spectra of the trion emission as a function of frequency $f$ for a driving amplitude $V_0$ = 100 mV. (b) Frequency-resolved PL spectra of trion emission with different tensile stresses. From left to right, the DC voltages were 0, 10, 20, 30, 40, and 50 V. The AC driving amplitude was kept at 25 mV. (c) Resonance $f_0$ and corresponding Q factor versus DC voltage applied to the actuator fingers (axial stress).

Although oscillations are often driven and detected using laser radiation[19], we wish to verify whether it is possible to use our (bulky) PMN-PT actuator to drive oscillations at frequencies closer to the recombination rate of our emitters (~1 GHz). Here, we explore the presence of possible resonances on another membrane with GaAs QDs up to ~100 MHz using a square-wave function for $V_{AC}$. To study the oscillations in time-domain we performed series of time-resolved PL measurements using an avalanche photodiode (APD)[39]. The general measurement procedures are as follows: first we choose a mechanical resonance, indicated by line broadening in time-integrated PL (here we also chose a trion emission line), then a monochromator is used to select emission at several particular wavelengths and the photons are detected by an APD by the time correlated counting electronics triggered by the function generator. Lastly, we collect all the data (three-dimensional matrix of wavelength-time-photon count) and plot the color-coded time-resolved (phase-locked) QD emission shift.

The result for a driving frequency of 19.75 MHz is shown in Fig. 4 (a), where the plateau is attributed to the used square-wave. The maximum resonant frequency we observed was around 61.5 MHz, see Fig. 4(b). We note that in this experiment the rate at which the emission energy shifts exceed values of 50 μeV/ns. Taking into account that the typical lifetime of the excitonic transitions in the state-of-the-art GaAs QDs is of the order of 250 ps and the corresponding natural linewidth is 2.3 μeV, this means that the strain modulation produces energy shifts of the order of 5 times the natural linewidth during the lifetime of the exciton. As strain also changes the excitonic fine-structure splitting[40,41], this speed may be already sufficient to observe non-trivial phenomena, such as Landau-Zener transitions[42,43]. Another application scenario is to use mechanical oscillations to control the state of a single spin confined in a QD via g-tensor modulation techniques at constant magnetic fields[44]. In fact, it has been recently found out that the g-tensor of holes confined in QDs reacts sensitively to strain[19,45]. Compared to electric-field based g-tensor modulation, a strain-field g-tensor modulation may be beneficial because of reduced hole dephasing produced by charge noise[46].

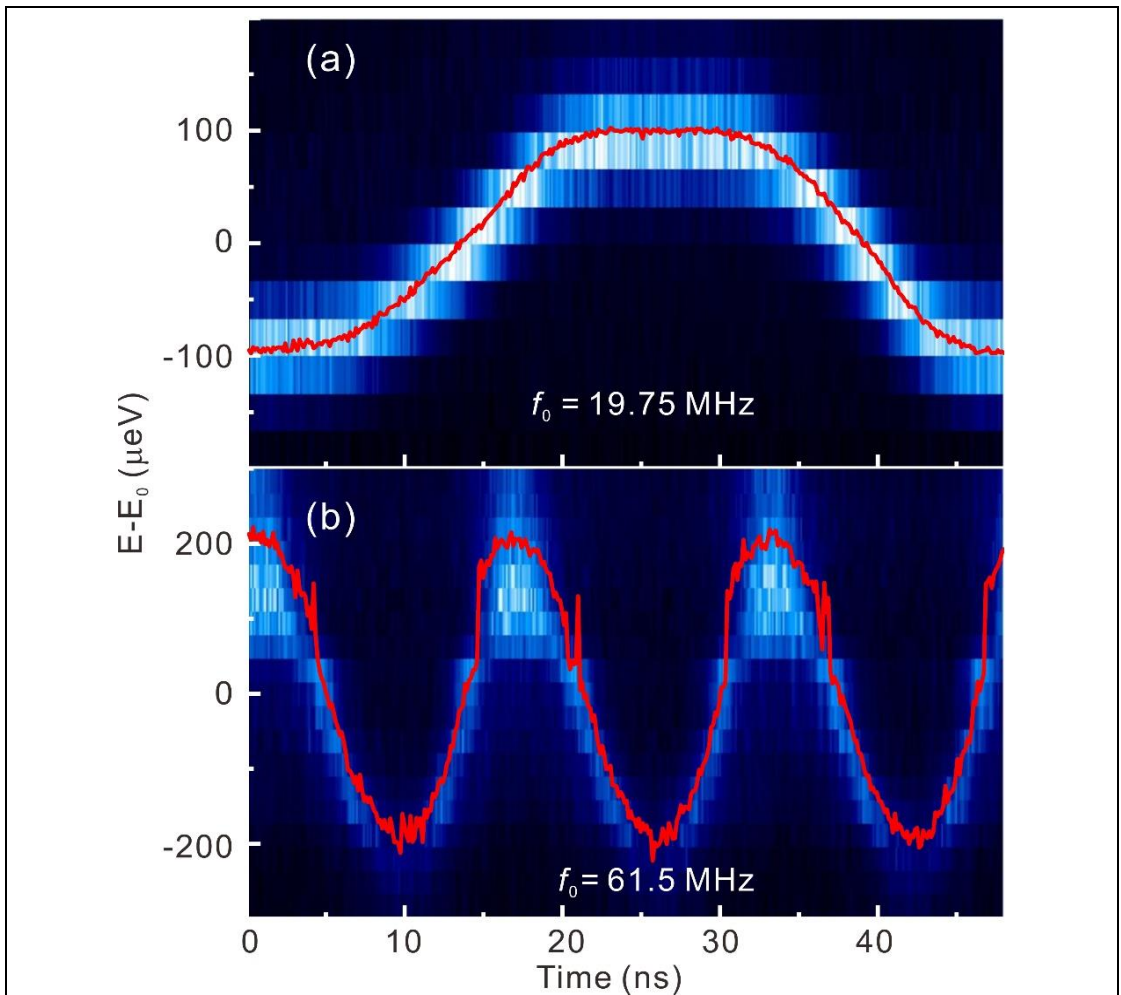

Figure 4. Time-resolved PL evolution of the exciton emission of a GaAs QD driven with a square wave with resonance frequency $f_0$ at (a) 19.75 MHz and (b) 61.5 MHz with a driving voltage $V_0$=1.9 V and DC voltage $V_{DC} = 0$. The red solid lines are the peak position extracted from Lorentz fits of the PL emission at fixed time.

In conclusion, we have investigated here the possibility of using micromachined piezoelectric actuators to tune the frequency of oscillations in micromechanical resonators consisting of semiconductor nanomembranes with embedded QDs. For a proof-of-principle we have used simple nanomembrane beam-oscillators directly bonded onto the actuator, which we use both to drive vibrations and control their resonant frequency. If required, tuning ranges substantially higher than those reported here are in principle possible. To this end, we envision having the mechanical resonators suspended on a well-defined rigid frame bonded to the PMN-PT actuator. This approach would open up the possibility to tune the resonant frequency of state-of-the-art optomechanical resonators, such as those based on silicon-nitride[47]. Due to the possibility of operating the actuators in vacuum and low temperature, the approach is more flexible than cantilever-bending method and may be a useful alternative to on-chip electrostatic actuation for fine-control of the resonance frequency of advanced optomechanical resonators operated in the quantum regime.

**Acknowledgement**

This work was supported by the FWF (P 29603), the Linz Institute of Technology (LIT), the LIT Secure and Correct Systems Lab funded by the state of Upper Austria, the EU project HANAS (No. 601126210), AWS Austria Wirtschaftsservice (PRIZE Programme, Grant No. P1308457), the European Research council (ERC) under the European Union's Horizon 2020 Research and Innovation Programme (SPQRel, Grant agreement No. 679183) and by the German Excellence Initiative via the Cluster of Excellence Nanosystems Initiative Munich (NIM). X. Yuan acknowledges support by China Scholarship Council (CSC, No. 201306090010). Y. Huo thanks support by NSFC (No. 11774326) and STCSM (No. 17ZR1443900, &No. 17PJ1409900). J.M.-S.

acknowledges support through the Clarín Programme from the Government of the Principality of Asturias and a Marie Curie-COFUND European grant (PA-18-ACB17-29). The authors thank M. Reindl, D. Huber for help with the processing and optical characterization and A. Halilovic, A. Schwarz, S. Bräuer, U. Kainz and E. Vorhauer for the technical support.